\documentclass[%
 reprint,
 superscriptaddress,
 amsmath,amssymb,
 aps,
 prl
]{revtex4-2}

\usepackage{graphicx}% Include figure files
\usepackage{dcolumn}% Align table columns on decimal point
\usepackage{bm}% bold math
\usepackage{xcolor}
\usepackage{multirow}
\usepackage{enumerate}
\begin{document}

\preprint{APS/123-QED}

%\title{Sextupole Implementation of Dynamic Focusing in Crab Crossing}% Force line breaks with \\
\title{Learning Dynamic Aperture from One-turn Maps}

\author{Derong Xu}\email{dxu@bnl.gov}
\affiliation{Brookhaven National Laboratory}%Lines break automatically or can be forced with \\

\date{\today}% It is always \today, today,

\begin{abstract}
Dynamic aperture evaluation relies on long-term tracking, while existing machine-learning surrogates remain difficult to generalize across machines. We demonstrate that coarse-grained dynamic aperture can be learned directly from suitably encoded one-turn maps. By reformulating dynamic-aperture prediction as an image segmentation problem, a deep surrogate model captures the long-term stability topology and transfers to realistic multidimensional Electron-Ion Collider Electron Storage Ring tracking. Failure analysis identifies a challenging resonant regime in which invariant tori are strongly deformed yet remain unbroken. These results establish a proof-of-principle that practical surrogate models can be constructed from one-turn transport information.
\end{abstract}

\maketitle

%\section{Introduction}

Circular accelerators are essential tools for scientific discovery, ranging from synchrotron light sources to high energy colliders. 
The nonlinear magnetic fields required for chromatic correction render the beam dynamics nonlinear, while stable operation requires charged particles to remain confined over billions of turns. 
Dynamic aperture (DA) is therefore a central figure of merit for nonlinear beam dynamics. 
In this work, DA is defined as the set of surviving initial conditions sampled within a finite phase space region at prescribed resolution after tracking for a fixed number of turns. 
Conventional DA is often characterized by the connected stable region surrounding the stable fixed point. 
However, disconnected stable islands reflect the phase-space structure and should be reproduced by a predictive model.

Artificial intelligence for dynamic aperture studies (AI4DA) has recently been developed. One approach predicts long-term particle stability from short-term trajectories, enabling fast DA evaluation for a given lattice \cite{Wan_2022}. Other studies train deep surrogate models with machine parameters as inputs, including tune, chromaticity, octupole strength, and magnetic error seeds, to predict DA over many configurations of a fixed accelerator \cite{instruments8040050}. Surrogate models have also been incorporated into DA optimization, where neural networks, Gaussian processes, active learning, and Bayesian algorithm execution reduce the number of expensive tracking evaluations \cite{PhysRevAccelBeams.24.014601,Zhang2025MultipointBAX}. These studies demonstrate substantial speedup, but also expose a common limitation: surrogate accuracy relies on sufficient coverage of the relevant parameter space, and practical optimization often requires retraining or active data acquisition. Existing AI4DA approaches have so far been demonstrated primarily within fixed lattices or restricted configuration families.

\begin{figure}
    \centering
    \includegraphics[width=0.99\columnwidth]{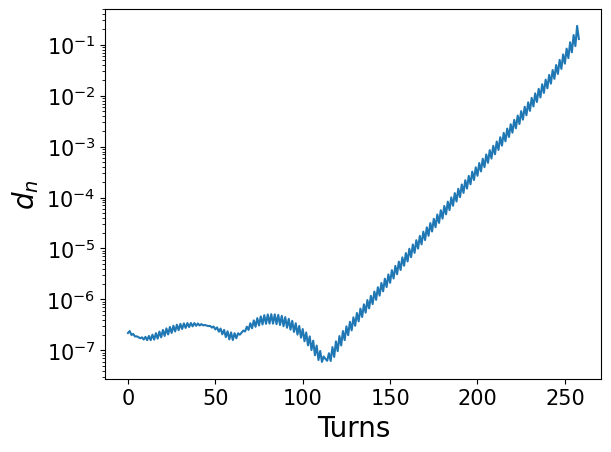}
    \caption{Phase space distance 
    $d_n=\sqrt{\left(x_{n,1}-x_{n,2}\right)^2+\left(p_{n,1}-p_{n,2}\right)^2}$
 between two nearby trajectories near the dynamic aperture boundary.
 The two trajectories remain close during the initial turns,
 while one particle survives after $16,000$ turns and the other is lost after $255$ turns.}
    \label{fig:trajectoryExample}
\end{figure}

%To generalize DA prediction across different machines and lattices, one possible route is to infer single particle stability from limited time series data. 
One possible route toward machine-independent stability prediction is to infer long-term particle survival from limited time series trajectory.
Related studies on regular and chaotic classification have demonstrated that deep neural networks can identify chaotic features from trajectories and exhibit nontrivial transfer capability across dynamical systems \cite{BOULLE2020132261,Celletti2022}. However, DA prediction differs fundamentally from generic chaos classification. Near the DA boundary, nearby initial conditions can exhibit nearly indistinguishable early trajectories while having different survival outcomes after long term tracking. A representative example in 2D phase space is shown in Fig.~\ref{fig:trajectoryExample}. In addition, nonlinear phase space transport can generate fine fractal structures near the stability boundary \cite{RevModPhys.81.333}, rendering the microscopic DA boundary highly resolution dependent. A prescribed resolution therefore becomes essential to determine the DA boundary.
%, defining a coarse grained DA as the learnable representation of the phase space structure.

For circular accelerators, single particle dynamics is completely determined by the one-turn map. By ``one-turn,'' we refer to the standard Poincar\'e map that advances the phase-space coordinates by one period of the system. The one-turn map is obtained by integrating the equations of motion over one revolution. Assuming a periodic Hamiltonian satisfying $H(s)=H(s+C)$, where $C$ is the ring circumference, the one-turn map is defined as
\begin{equation}
\mathcal{M}
=
\mathcal{S}
\exp\!\left(
\int_0^C ds\, : -H(s):
\right),
\end{equation}
where $\mathcal{S}$ denotes path ordering along the longitudinal coordinate $s$. The notation $:H:$ denotes the Lie operator acting through the Poisson bracket, $:H:g=[H,g]$, following the formalism of A. J. Dragt \cite{AJDragt}. Since dynamic aperture arises from repeated iteration of the one-turn map, it is fully determined by $\mathcal{M}$. The lattice configuration parameters, including magnet strengths, lengths, and element types, are generally highly redundant from the viewpoint of nonlinear dynamics. For example, a well optimized lattice may exhibit dynamics close to a weakly perturbed linear oscillator, characterized by relatively few effective dynamical degrees of freedom despite a high dimensional lattice parameter space.

The one-turn symplectic map admits an exact Dragt–Finn factorization into an infinite series of Lie transformations \cite{dragt1976lie},
\begin{equation}
\mathcal{M}
=
\exp(:f_2:)
\exp(:f_3:)
\exp(:f_4:)
\cdots ,
\label{eq:dragtFinn}
\end{equation}
where $f_n$ denotes a homogeneous polynomial generator of order $n$. After normalization by the linear optics, the quadratic generator reduces to a rotation
in phase space,
\begin{equation}
f_2
=
-\frac{\mu}{2}(x^2+p^2).
\end{equation}
The higher-order generators encode the nonlinear transport structure of the map.

%In realistic synchrotrons, the dominant nonlinearities typically originate from low-order magnetic elements, including quadrupoles, sextupoles, and octupoles, while higher-order generators mainly arise from composition of lower-order elements rather than standalone devices. Consider a phase-space neighborhood characterized by a typical amplitude scale $A$. For an $n$th-order homogeneous generator $f_n$, the nonlinear contribution scales as $A^n$, while the linear dynamics scales as $A^2$. Therefore, relative to the linear term, the effective strength of higher-order generators scales as $A^{n-2}$, and higher-order nonlinearities become progressively weaker toward the stable fixed point. Over sufficiently long tracking time, even weak nonlinear terms can influence the dynamic aperture through cumulative resonance effects. However, within a sufficiently small observation window and finite tracking horizon, high-order generators beyond a certain order may remain close to the identity transformation and have limited impact on the binary survival label. This heuristic is consistent with the accelerator physics rule of thumb that higher-order resonances are generally less dangerous. Coarse resolution, finite observation window, and finite tracking time are therefore essential to render the binary survival problem well posed for machine learning.

In a phase-space neighborhood of characteristic amplitude $A$, an $n$th-order homogeneous generator $f_n$ contributes at order $A^n$, whereas the linear dynamics scales as $A^2$. The relative strength of higher-order generators therefore scales as $A^{n-2}$, implying that sufficiently high-order nonlinearities are increasingly suppressed at small amplitudes. While such terms can influence dynamic aperture through long-term resonance accumulation, their contribution can be controlled by restricting the characteristic phase-space scale and tracking horizon. Coarse resolution, finite observation windows, and finite tracking horizons therefore motivate the survival prediction a practically learnable target.

%Motivated by these considerations, 
This Letter adopts a truncated Dragt-Finn representation up to seventh order to generate a large ensemble of synthetic one-turn maps in two-dimensional normalized phase space. For each map, one-turn transport is sampled on a prescribed phase-space grid and used as the surrogate input. 
The central question is whether finite-horizon dynamic aperture can be inferred from one-turn transport information. 
The representation contains only limited information sampled on a finite grid and therefore does not uniquely determine the exact map.
To investigate this question, particles are tracked for $1024$ turns (see Appendix A) to generate binary survival labels, with the dynamic aperture defined by the collection of surviving pixels. A surrogate model is then trained to predict the survival label of each phase-space pixel.

%The dynamic aperture prediction problem is reformulated as an image segmentation task on a fixed phase-space grid. Each generated map is represented by a $64\times64\times C$ tensor, with channel dimension
%\[
%C=(x_0,p_0,x_1,p_1,\mathrm{mask}),
%\]
%where $(x_0,p_0)$ denote the initial coordinates on the phase-space grid, $(x_1,p_1)$ denote their one-turn image under the map, and $\mathrm{mask}$ indicates whether the one-turn tracking result is valid. 
%For strongly nonlinear maps, some initial conditions may fail to admit a valid one-turn image and are excluded through the mask channel.
%The pairwise representation $(x_0,p_0)\mapsto(x_1,p_1)$ provides a high-resolution local sampling of the one-turn map on the prescribed grid. The network output is a binary segmentation map labeling each grid point as alive or lost after $1024$ turns tracking.

%This formulation translates dynamic aperture prediction into a dense pixel-wise classification problem analogous to classical image segmentation. The local one-turn transport structure is encoded through $(x_0,p_0,x_1,p_1)$, while disconnected stable islands naturally appear as segmentation targets.
%Image segmentation is a mature machine-learning paradigm with well-developed architectures and training methodologies, particularly in applications such as medical image analysis \cite{Litjens2017}.

The sampled one-turn transport is represented on a fixed phase-space grid through the local transport pairs $(x_0,p_0)\mapsto(x_1,p_1)$ together with a validity mask, thereby reformulating dynamic aperture prediction as an image segmentation problem. In this representation, disconnected stable islands naturally appear as segmentation targets. This formulation enables the use of mature image segmentation architectures originally developed for tasks such as medical image analysis \cite{Litjens2017}.

%The truncated Dragt–Finn representation used in this work contains $31$ free parameters. For a given nonlinear map, an appropriate observation window can always be chosen such that the nonlinear coefficients remain of order unity. The phase-space sampling window is therefore normalized to $(x_0,p_0)\in[-1,1]^2$.

%The construction of a suitable training dataset is nontrivial. A natural first strategy is to sample the tune uniformly from $(0,1)$ and the remaining nonlinear coefficients uniformly from $(-1,1)$. In principle, given sufficient samples, the random sampling can cover the selected representation space. In practice, exhaustive coverage is computationally prohibitive. The resulting dataset is strongly biased toward highly nonlinear maps with small stable regions, leading to poor model performance for weakly nonlinear cases.

%To broaden the coverage of nonlinear regimes, a final dataset is constructed through multi-scale sampling of the observation window (see Appendix~B). 
%Figure~\ref{fig:samplingHistogram} compares the distribution of stable pixel fractions for direct sampling and multi-scale sampling. Multi-scale sampling substantially broadens the distribution, covering both weakly nonlinear cases with large stable fractions and strongly nonlinear cases with fragmented small-aperture structures. 
%Such broad coverage is essential for constructing a surrogate model applicable to realistic dynamic aperture optimization problems, where optimization trajectories may traverse both weak and strong nonlinear regimes.

The observation window is chosen so that the nonlinear generator coefficients remain bounded by order unity, while the corresponding phase-space coordinates $(x_0,p_0)$ are normalized to $[-1,1]^2$. Direct random sampling produces datasets strongly biased toward maps with low stable-pixel fractions. To broaden the coverage of nonlinear regimes, the final dataset is generated using multi-scale sampling of the observation window (see Appendix~B). As shown in Fig.~\ref{fig:samplingHistogram}, this procedure substantially broadens the distribution of stable pixel fractions, covering both weakly nonlinear maps with large stable regions and strongly nonlinear maps with fragmented small-aperture structures. The resulting dataset contains two million samples. The train, validation, and test sets are generated from disjoint sets of base maps, and all multi-scale samples derived from the same base map remain in the same split. The validation and test sets each contain $10^4$ samples.

\begin{figure}
    \centering
    \includegraphics[width=0.99\columnwidth]{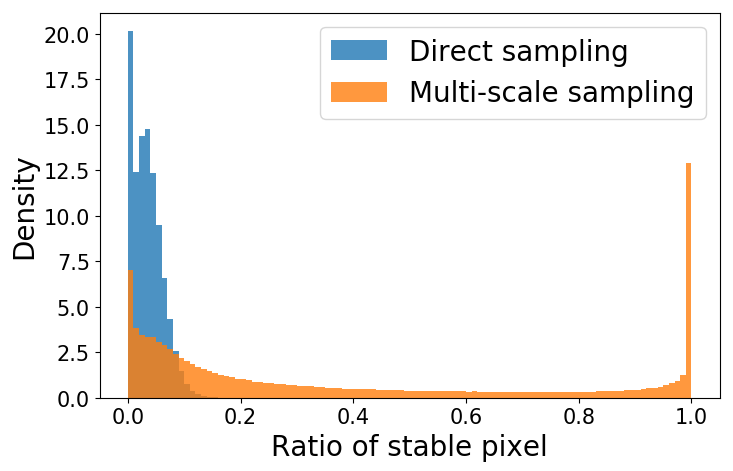}
    \caption{Histogram of stable pixel fractions. Direct sampling yields an average stable fraction of $3.5\%$, whereas multi-scale sampling increases the average to $38\%$ and substantially broadens the coverage across nonlinear regimes.}
    \label{fig:samplingHistogram}
\end{figure}

%The multi-scale construction yields approximately two million map samples, which are separated into independent training, validation, and test subsets containing approximately $1.98\times10^6$, $10^4$, and $10^4$ samples, respectively.

%Following the image segmentation formulation introduced above, the surrogate model is constructed using convolutional neural networks (CNNs), which learn spatial patterns through local filters and hierarchical feature extraction \cite{LeCun2015}. As a baseline, we adopt the U-Net architecture \cite{Ronneberger2015}, an encoder-decoder network originally developed for image segmentation. The encoder progressively compresses the input representation to extract multi-scale features, while the decoder reconstructs the pixel-wise output using skip connections that preserve fine spatial structures.

Following the image segmentation formulation introduced above, the surrogate model is constructed using convolutional neural networks (CNNs) \cite{LeCun2015}. As a baseline, we adopt the U-Net architecture \cite{Ronneberger2015}, a standard encoder-decoder network for image segmentation.

Since dynamic aperture may contain disconnected islands and long-range phase-space correlations, purely local feature extraction may be insufficient. We therefore additionally investigate an attention U-Net architecture \cite{Oktay2018}, in which attention blocks dynamically reweight feature maps according to their relevance. Model performance is evaluated using the intersection-over-union (IoU) metric,
\begin{equation}
\mathrm{IoU}
=
\frac{\mathrm{card}(Y_{\rm pred}\cap Y_{\rm true})}
{\mathrm{card}(Y_{\rm pred}\cup Y_{\rm true})},
\end{equation}
where $Y_{\rm pred}$ and $Y_{\rm true}$ denote the predicted and true surviving-pixel regions. Larger IoU values indicate better agreement.

%Since the dynamic aperture may contain disconnected islands and long-range phase-space correlations, purely local feature extraction may be insufficient. We therefore additionally investigate an attention U-Net architecture \cite{Oktay2018}, in which attention blocks dynamically reweight feature maps according to their global relevance. Model performance is evaluated using the intersection-over-union (IoU) metric, 
%\begin{equation}
%\mathrm{IoU}=
%\frac{\mathrm{card}\!\left(Y_{\rm pred}\cap Y_{\rm true}\right)}
%{\mathrm{card}\!\left(Y_{\rm pred}\cup Y_{\rm true}\right)},
%\end{equation}
%which is widely used in image segmentation to quantify the overlap between the predicted and true surviving-pixel regions. An IoU approaching unity indicates accurate reconstruction, whereas an IoU near zero corresponds to poor agreement.

\begin{figure}
    \centering
    \includegraphics[width=0.99\columnwidth]{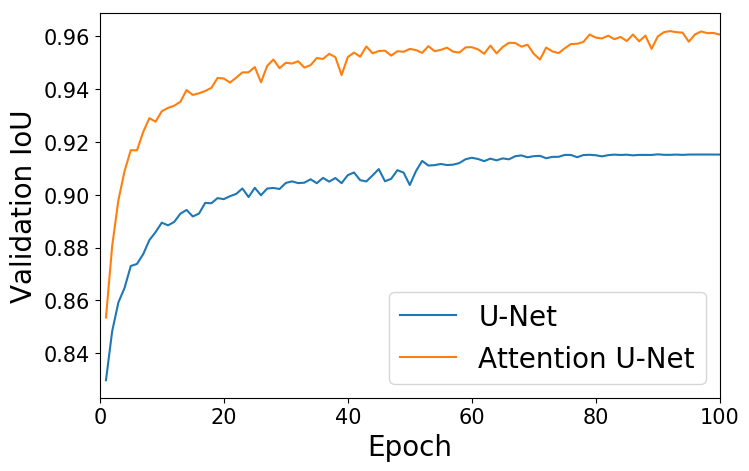}
    \caption{Validation IoU versus training epoch for the U-Net and attention U-Net architectures. Attention blocks substantially improve the prediction accuracy.}
    \label{fig:trainingIoU}
\end{figure}

Figure~\ref{fig:trainingIoU} compares the validation IoU during training for the two architectures. The baseline U-Net reaches a validation IoU of approximately $0.915$, whereas the attention U-Net achieves approximately $0.962$.
%The high IoU indicates that coarse dynamic-aperture topology can be reconstructed accurately for most sampled one-turn maps.
An IoU approaching unity indicates that most of the stable-region topology can be inferred from one-turn transport information alone.
The training results demonstrate that finite-resolution samples of the one-turn transport contain sufficient information to determine the corresponding finite-horizon dynamic aperture.

The substantial improvement produced by attention carries a direct physical implication. Dynamic-aperture topology cannot be inferred solely from the local one-turn transport surrounding an individual phase-space point. Instead, accurate prediction requires information about larger-scale phase-space structures, including disconnected stable islands and extended transport patterns. 
The attention gain therefore provides evidence that global information contributes to dynamic-aperture inference. 

%To probe transferability beyond the training distribution, we apply the surrogate to realistic accelerator tracking. The Electron-Ion Collider (EIC) is a next-generation collider designed to use electrons as probes to explore the structure of nucleons and nuclei \cite{osti_1765663}. The EIC adopts a swap-out electron injection scheme to minimize perturbations to the hadron beam. To support its scientific goals, the EIC Electron Storage Ring (ESR) is designed to achieve approximately $10\sigma$ dynamic and momentum aperture \cite{PhysRevAccelBeams.25.071001}. 

To probe transferability beyond the training distribution, we apply the surrogate to realistic accelerator tracking. The Electron-Ion Collider (EIC) is a next-generation collider currently under construction \cite{osti_1765663}. Its Electron Storage Ring (ESR) is designed to achieve approximately $10\sigma$ dynamic and momentum aperture \cite{PhysRevAccelBeams.25.071001}, providing a realistic test of the surrogate beyond the synthetic training ensemble.

We perform $1024$-turn ESR tracking using XSuite \cite{Iadarola:2023fuk} in a realistic multidimensional lattice model with radiation damping and RF disabled at $\delta=0$. To apply the surrogate, the initial vertical coordinates are fixed at $y=0$ and $p_y=0$, while $(x,p_x)$ are uniformly sampled within $\pm24\sigma_x$ and $\pm24\sigma_{p_x}$, respectively. The sampled coordinates are normalized by the corresponding $24\sigma$ ranges and used as surrogate inputs without retraining the model.

Zero initial vertical amplitudes do not freeze the additional degrees of freedom. Through nonlinear coupling and multidimensional transport, particles can still evolve in the vertical phase space. Quantitative discrepancies between the surrogate prediction and realistic tracking are therefore anticipated.

\begin{figure}
    \centering
    \includegraphics[width=0.99\columnwidth]{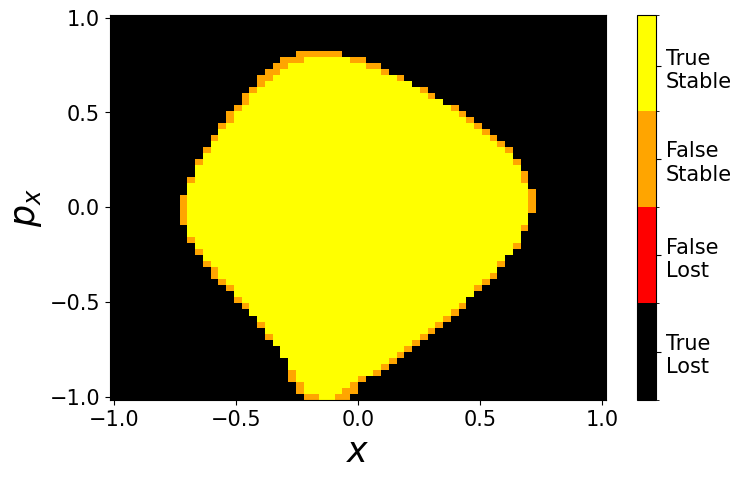}
    \caption{Out-of-distribution application of the surrogate model to realistic EIC ESR tracking. Although trained only on synthetic two-dimensional truncated Dragt--Finn maps, the surrogate reproduces the coarse ESR dynamic aperture obtained from multidimensional XSuite tracking. Colors denote true stable, true lost, false stable, and false lost pixels.}
    \label{fig:esrPrediction}
\end{figure}

Nevertheless, Fig.~\ref{fig:esrPrediction} shows that the surrogate reproduces the coarse dynamic aperture remarkably well. The prediction captures the large connected stable region, the global asymmetric shape, and characteristic boundary structures, including the flattened upper boundary and narrow lower cusp. Most discrepancies are localized near the dynamic aperture boundary and are dominated by false-stable pixels, while false-lost pixels remain rare. Given the substantial mismatch between the synthetic two-dimensional training maps and realistic multidimensional ESR tracking, this level of agreement suggests that the surrogate has learned transferable features of the underlying nonlinear dynamics.

\begin{figure}
    \centering
    \includegraphics[width=0.99\columnwidth]{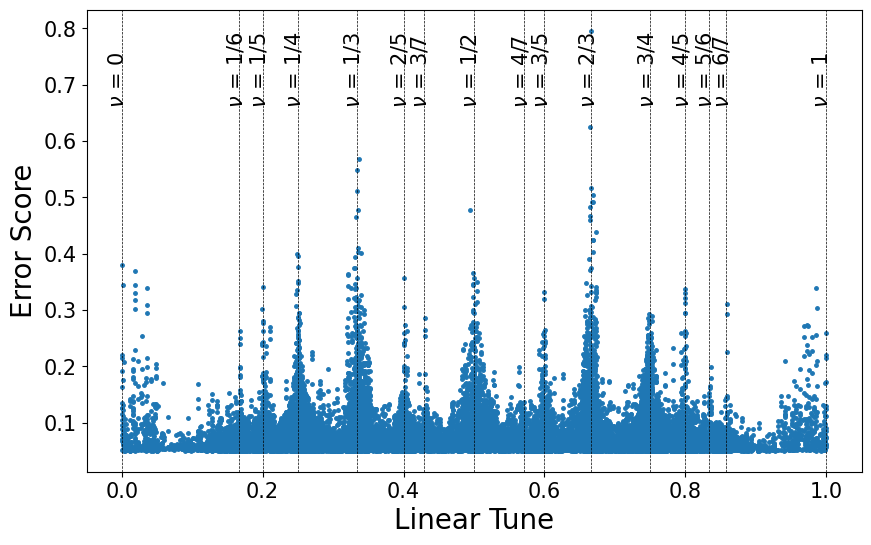}
    \caption{
        Prediction error score versus linear tune for all training samples with score greater than $0.05$, comprising $25,655$ cases out of approximately $1.98\times10^6$ training samples.}
    \label{fig:trainingErrorScore}
\end{figure}

Despite the strong average performance and successful ESR application, catastrophic failures still occur in a small subset of cases. Defining the error score as the fraction of misclassified pixels, Fig.~\ref{fig:trainingErrorScore} shows that the largest errors cluster near low-order resonance lines. The most severe failures occur when resonance proximity is combined with small observation windows. In these cases, direct tracking shows that nearly all pixels survive within the selected window, whereas the surrogate predicts that only a small fraction remain stable.

Inspection of representative examples reveals that the invariant tori remain intact despite strong resonant distortion (see Appendix~C). The surrogate correctly identifies the nearby resonance, but systematically overestimates its impact on long-term stability. Detecting resonance therefore appears easier than determining whether resonance has actually destroyed transport barriers. %, suggesting that resonance-induced topology deformation is more difficult to learn than topology destruction.

One possible explanation for these failures is insufficient exposure to rare resonant topologies in the training data. Exact resonance conditions occupy a small measure in parameter space, making strongly distorted yet stable phase-space structures statistically rare. To test this hypothesis, we continue training with three additional independently generated datasets, each containing $2\times10^6$ samples.

%Figure~\ref{fig:tailImprovement} shows the tail statistics evaluated on the same held-out test set. Continued training systematically reduces prediction errors across all thresholds, from moderate errors (error score $>0.05$) to severe failures (error score $>0.5$). 
%The persistence of a residual high-error population, however, further demonstrates  that highly resonant yet unbroken invariant structures remain intrinsically challenging for the current surrogate.

\begin{figure}
    \centering
    \includegraphics[width=0.99\columnwidth]{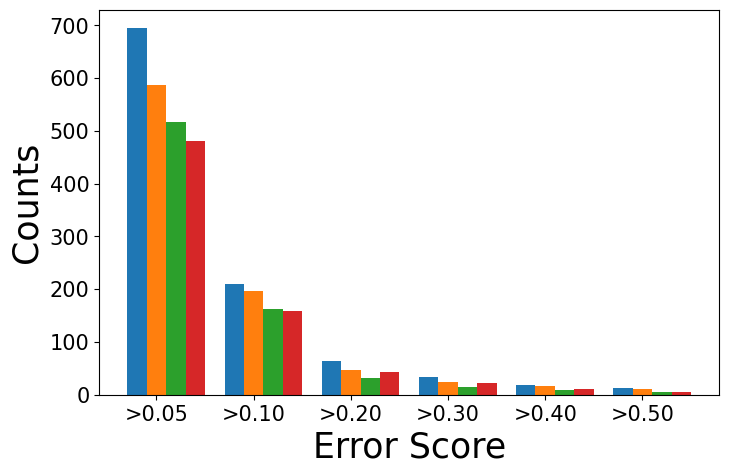}
    \caption{Tail statistics of the prediction error evaluated on a fixed test set ($10^4$ samples) after 
    successive training rounds with three additional datasets (orange, green and red bars). Bars show the number of test samples with error score 
    exceeding each threshold.}
    \label{fig:tailImprovement}
\end{figure}

Figure~\ref{fig:tailImprovement} shows the tail statistics evaluated on the same held-out test set. Continued training reduces prediction errors across a broad range of thresholds, indicating that additional data improves difficult cases. However, the improvement is not uniform across the tail: moderate and intermediate errors continue to decrease, whereas the severe failures saturate after the third training stage. This behavior suggests that the residual catastrophic failures are not simply removed by adding more training data, but are associated with a distinct and substantially more challenging dynamical regime.

This failure mode has a natural interpretation from the viewpoint of KAM theory \cite{dumas2014kam}. While KAM theory guarantees the persistence of invariant tori under sufficiently weak perturbations, it provides little practical guidance on whether a given resonant perturbation is sufficiently weak. The concentration of residual errors suggests that these cases lie precisely in this delicate regime, where resonance signatures are evident but the associated invariant tori still survive. 
%Since the surrogate receives only finite-resolution one-turn transport information and provides only an approximate representation of the dynamics, it is not surprising that this regime remains particularly challenging. 
%The residual errors therefore appear to be concentrated near the boundary between topology deformation and topology destruction.

%Although these failures reveal a genuine limitation of the current surrogate, they do not substantially weaken its utility for accelerator design studies. Only approximately $5\%$ of the validation and test samples exhibit an error score exceeding $5\%$, implying that the surrogate prediction is accurate at the few-percent level for the majority of cases. Such accuracy is sufficient for practical machine optimization, where the objective is typically to identify promising candidate solutions rather than the exact global optimum. Moreover, realistic accelerator operation usually avoids working points near low-order resonance lines because of their well-known detrimental effects on beam dynamics.

Although these failures reveal a genuine limitation of the current surrogate, they remain confined to a small fraction of cases. Only approximately $5\%$ of the validation and test samples exhibit an error score exceeding $5\%$, indicating that the surrogate reproduces the dynamic-aperture topology at the few-percent level for the vast majority of cases. Such accuracy is likely sufficient for many accelerator design and optimization studies, where the primary objective is often to identify promising regions of parameter space rather than the exact optimum. Furthermore, the observed failures are concentrated near low-order resonances, which are generally avoided in practical accelerator design because of their adverse impact on beam dynamics.

Because catastrophic failures cannot be entirely eliminated within the current framework, practical deployment would benefit from diagnostics for unreliable predictions. Two candidate diagnostics are available. First, the network architecture can include an auxiliary head that predicts the stable-pixel fraction, whose inconsistency with the pixel-wise segmentation output can signal potential prediction failures. Second, the input representation possesses an exact coordinate-scaling redundancy: rescaling the coordinate channels changes only the coordinate units while leaving the underlying one-turn map and survival labels unchanged. Sensitivity of the surrogate prediction under such transformations can therefore provide an additional diagnostic probe. A scaling augmentation study is presented in Appendix~D.

%In summary, this work demonstrates that dynamic aperture can be learned from suitably encoded one-turn maps. The resulting surrogate captures long-term stability topology, including disconnected stable structures and global nonlinear transport patterns, and transfers to realistic EIC ESR tracking beyond the synthetic training distribution. The failure analysis further clarifies the scope of learnability: catastrophic failures concentrate near resonant regimes, where the surrogate correctly identifies nearby resonances but misjudges whether the associated transport barriers remain intact. Continued improvement with additional training data suggests that these failures are partly associated with insufficient exposure to rare resonant topologies, while the residual hard cases define a challenging regime of nonlinear beam dynamics. 
%Taken together, these results establish a proof-of-principle that practical surrogate models can be constructed from one-turn map encodings,
%which retains sufficient information about long-term stability for accelerator design.
%More broadly, the present work suggests that one-turn transport sampled on finite grids retains substantial information about long-term stability that is not typically exploited in accelerator design.

In summary, this work demonstrates that dynamic aperture can be learned from suitably encoded one-turn maps. The resulting surrogate transfers to realistic EIC ESR tracking beyond the synthetic training distribution. The observed failure modes suggest that learnability is not uniform across phase space, with the most challenging cases concentrated near resonant structures where topology deformation and topology destruction become difficult to distinguish. These results establish a proof-of-principle that finite-resolution one-turn transport information retains sufficient information to infer coarse-grained long-term stability, opening a path toward practical surrogate models for accelerator design and optimization.

%provides a path toward machine-independent DA prediction for circular accelerators.

%How to encode nonlinear dynamics into a learnable representation

%Trivial result: mask overlap with label too many, the network may not learning dynamics if pre-track too many turns. A strong baseline could be 100 turn label
%used as proxy for 1000 turn label

%the learnable object is the coarse-grained DA structure rather than the exact microscopic boundary

%A. Hénon map benchmark
%B. MLP point-wise classifier as a minimal predictor
%C. Difference map: errors localize near the boundary
%D. Lyapunov argument: finite-time trajectories cannot distinguish nearby initial conditions
%E. Conclusion: the learnable object is the coarse-grained DA structure

%(3) engineering effort for a general Dragt-Finn representation, 
%(4) summay part

%%%%%%%%%%%%%%%%%%%%%%%%%%%%%%%%%%%%%%%%%%%%%%%%%%%%%%%%%%%%%%%%%%%%%%%%%%%%%%%%%%%%%%%%%%%%%%%%%%%%%%%%%

\begin{acknowledgments}
This work was supported by Brookhaven Science Associates, LLC under Contract No. DE-SC0012704 with the U.S. Department of Energy, and by a U.S. Department of Energy Early Career Award.
This research used resources of the National Energy Research Scientific Computing Center (NERSC), a U.S. Department of Energy Office of Science User Facility operated under Contract No. DE-AC02-05CH11231.

\end{acknowledgments}
%\nocite{*}

%%%%%%%%%%%%%%%%%%%%%%%%%%%%%%%%%%%%%%%%%%%%%%%%%%%%%%%%%%%%%%%%%%%%%%%%%%%%%%%%%%%%%%%%%%%%%%%%%%%%%%%%%
\appendix

\section{Appendix A: Truncated Dragt-Finn maps tracking}
\label{app:integration}

The one-turn map in Eq.~(\ref{eq:dragtFinn}) is evaluated through symplectic integration of the Lie generators. The linear part corresponds to a phase-space rotation after normalization by the Twiss functions. The nonlinear generators $f_n$ with $n\geq 3$ are evaluated numerically.

For strongly nonlinear generators with large coefficients, slicing is employed by replacing
\begin{equation}
f_n\rightarrow f_n/N_s
\end{equation}
and applying the corresponding map $N_s$ times to control integration error.

Each homogeneous polynomial is decomposed into monomial generators,
\begin{equation}
f_n=\sum_i f_{n,i}.
\end{equation}
Closed-form expressions for monomial generators can be found in Ref.~\cite{gjaja1994monomial}. 

The monomial maps are composed using Strang operator splitting \cite{Strang1968} to obtain a second-order approximation,
\begin{equation}
\mathcal{M}_2=
\prod_i
\exp\left(:\frac{f_{n,i}}{2}:\right)
\prod_i^{\leftarrow}
\exp\left(:\frac{f_{n,i}}{2}:\right),
\end{equation}
where the second product denotes reverse ordering. 

A fourth-order integrator is then constructed using Yoshida composition \cite{Yoshida1990},
\begin{equation}
\mathcal{M}_4=
\mathcal{M}_2(\alpha)
\mathcal{M}_2(\beta)
\mathcal{M}_2(\alpha),
\end{equation}
with
\begin{equation}
\alpha=\frac{1}{2-2^{1/3}},
\qquad
\beta=-\frac{2^{1/3}}{2-2^{1/3}}.
\end{equation}
Here $\mathcal{M}_2(\alpha)$ denotes the second-order map with all monomial generators scaled by $\alpha$.

The tracking scheme preserves the symplectic structure of the truncated one-turn map and remains stable across the generated map ensemble.

%%%%%%%%%%%%%%%%%%%%%%%%%%%%%%%%%%%%%%%%%%%%%%%%%%%%%%%%%%%%%%%%%%%%%%%%%%%%%%%%%%%%%%%%%%%%%%%%%%%%%%%%%
\section{Appendix B: Multi-scale dataset construction}
\label{app:dataset}

The final dataset is generated from randomly initialized truncated Dragt-Finn maps using multi-scale observation windows. The procedure is summarized below. 

\begin{center}
\begin{minipage}{0.95\columnwidth}
\hrule
\vspace{0.5em}
\textbf{Algorithm 1: Multi-scale construction of the training dataset}

\vspace{0.5em}
\textbf{Input:} grid size $N_g=64$, tracking horizon $N_{\rm turn}=1024$, slicing number $N_s=10$, window scale $s$, number of logarithmic bins $N_b=10$.

\vspace{0.5em}
\begin{enumerate}[(1)]
\item Generate a random truncated Dragt-Finn map $\mathcal M$.

\item Divide the window-scale interval into $N_b$ logarithmic bins.

\item For each bin, sample a scale factor $s$ uniformly within the bin.

\item For each grid point $(x_0,p_0)\in[-1,1]^2$, set the physical initial condition as
$(x,p)=s(x_0,p_0)$.

\item Track one turn to obtain $(x_1,p_1)=\mathcal M(x,p)$.

\item If $(x_1,p_1)$ is finite and satisfies $|x_1|,|p_1|<x_{\max}$, store
$(x_0,p_0,x_1/s,p_1/s,1)$ as the input feature. Otherwise, store
$(x_0,p_0,0,0,0)$.

\item Starting from $(x_1,p_1)$, iterate the map for $N_{\rm turn}$ turns.

\item Assign $Y(x_0,p_0)=1$ if all iterates remain finite; otherwise assign
$Y(x_0,p_0)=0$.

\item Save the feature tensor $X\in\mathbb{R}^{5\times N_g\times N_g}$ and the binary label map
$Y\in\{0,1\}^{N_g\times N_g}$.
\end{enumerate}

\vspace{0.3em}
\hrule
\end{minipage}
\end{center}

In this work, $x_{\max}=10$ is introduced to reject pathological one-turn images with excessively large coordinates (e.g. $|x_1|,|p_1|\gg1$) arising from strongly unstable maps. The mask channel records whether the one-turn image satisfies this validity criterion.
Each randomly generated map produces $N_b$ samples corresponding to different observation window scales, thereby broadening the one-turn map distribution across weakly and strongly nonlinear regimes.

%%%%%%%%%%%%%%%%%%%%%%%%%%%%%%%%%%%%%%%%%%%%%%%%%%%%%%%%%%%%%%%%%%%%%%%%%%%%%%%%%%%%%%%%%%%%%%%%%%%%%%%%%
\section{Appendix C: Catastrophic failure example}
\label{app:failureExample}
\begin{figure}
    \centering
    \includegraphics[width=0.99\columnwidth]{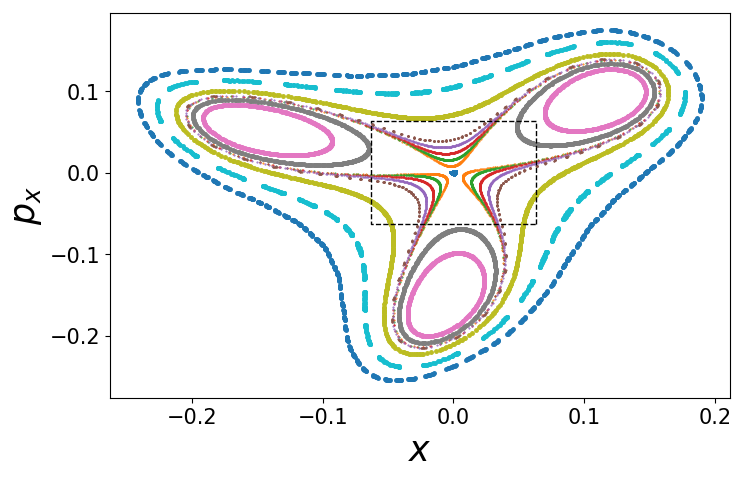}
    \caption{
    Representative catastrophic failure from the training dataset. The map lies close to the third-order resonance $\nu=1/3$. The invariant tori are strongly distorted, producing three large side islands and a compressed central region. The dashed box indicates the observation window. Although all pixels inside the window survive from direct tracking, the surrogate predicts most of them as lost.
    }
    \label{fig:failureExample}
\end{figure}

Figure~\ref{fig:failureExample} presents a representative catastrophic failure from the training dataset. The corresponding map has linear tune $\nu=0.33357$, close to the third-order resonance. The failure mechanism in this example is closely related to the mismatch between the global resonant topology and the local observation window. The tune proximity to $\nu=1/3$ produces a strongly distorted phase-space portrait with pronounced island structures and a compressed central region. However, the selected observation window lies inside a region where the invariant tori remain unbroken. Consequently, local stability is preserved despite the strong resonant signature visible in the full phase space. The surrogate correctly recognizes the nearby resonance, but appears overly sensitive to the associated topology deformation in this regime.

%%%%%%%%%%%%%%%%%%%%%%%%%%%%%%%%%%%%%%%%%%%%%%%%%%%%%%%%%%%%%%%%%%%%%%%%%%%%%%%%%%%%%%%%%%%%%%%%%%%%%%%%%
\section{Appendix D: Scaling symmetry and augmentation}
\label{app:scaling}

The input representation used in this work possesses a simple scaling redundancy. Given an input tensor
\[
(x_0,p_0,x_1,p_1,\mathrm{mask}),
\]
the transformed representation
\[
(sx_0,sp_0,sx_1,sp_1,\mathrm{mask})
\]
corresponds to the same one-turn map expressed in a different coordinate unit. The underlying nonlinear dynamics therefore remains unchanged, and the binary survival labels should be identical.

However, this invariance is not explicitly built into the network architecture. To investigate whether the surrogate can learn this symmetry, an additional training experiment is performed using scaling augmentation. During training, a random scaling factor
\[
s\sim U(0.8,1.2)
\]
is applied to the coordinate channels of each training sample, while the label map is unchanged.

\begin{figure}
    \centering
    \includegraphics[width=0.99\columnwidth]{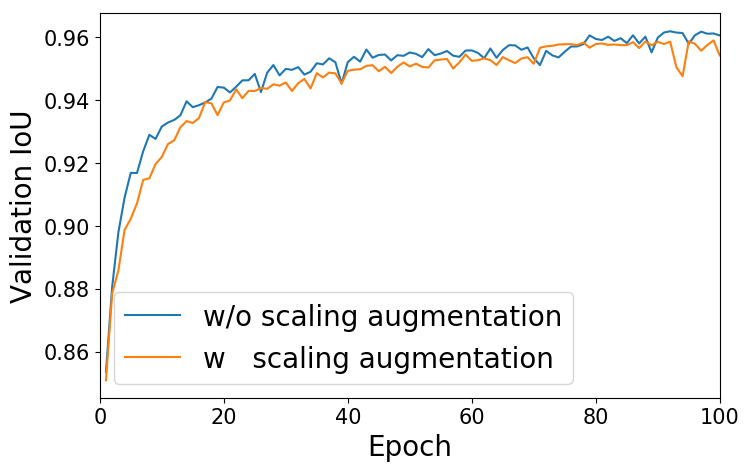}
    \caption{Validation IoU during training without and with scaling augmentation. Scaling augmentation enforces an exact coordinate redundancy of the input representation but yields prediction accuracy comparable to the baseline model. The scaling symmetry can nevertheless serve as a diagnostic probe for prediction robustness.}
    \label{fig:scalingAugmentation}
\end{figure}

Figure~\ref{fig:scalingAugmentation} compares the validation IoU obtained with and without scaling augmentation. The augmented model reaches a validation IoU comparable to the baseline model, with best performance approximately $0.960$ compared with $0.962$ without augmentation. The absence of significant improvement suggests that explicit enforcement of this symmetry is not essential for the present task.

Nevertheless, scaling invariance provides a useful diagnostic probe. Since the physical prediction should remain unchanged under coordinate rescaling, sensitivity of the surrogate output to such transformations may indicate regions where the prediction is unreliable. Although not a guaranteed failure detector, violations of approximate scaling consistency can help identify potential catastrophic failures.

%%%%%%%%%%%%%%%%%%%%%%%%%%%%%%%%%%%%%%%%%%%%%%%%%%%%%%%%%%%%%%%%%%%%%%%%%%%%%%%%%%%%%%%%%%%%%%%%%%%%%%%%%
\bibliography{ref}% Produces the bibliography via BibTeX.

\end{document}